# Evaluating Link-Based Techniques for Detecting Fake Pharmacy Websites

Ahmed Abbasi[1], Siddharth Kaza[2], Fatemeh "Mariam" Zahedi[1]

University of Wisconsin-Milwaukee[1], Towson University[2]

abbasi@uwm.edu, skaza@towson.edu, zahedi@uwm.edu

## Abstract

*Fake online pharmacies have become increasingly pervasive, constituting over 90% of online pharmacy websites. There is a need for fake website detection techniques capable of identifying fake online pharmacy websites with a high degree of accuracy. In this study, we compared several well-known link-based detection techniques on a large-scale test bed with the hyperlink graph encompassing over 80 million links between 15.5 million web pages, including 1.2 million known legitimate and fake pharmacy pages. We found that the QoC and QoL class propagation algorithms achieved an accuracy of over 90% on our dataset. The results revealed that algorithms that incorporate dual class propagation as well as inlink and outlink information, on page-level or site-level graphs, are better suited for detecting fake pharmacy websites. In addition, site-level analysis yielded significantly better results than page-level analysis for most algorithms evaluated.*

## 1. Introduction

Online pharmacies represent an invaluable source of information and products for consumers and individuals seeking health-related information. Accordingly, there has been considerable growth in the usage of Internet pharmacies (Easton, 2007; Wilford et al., 2006). Unfortunately, they have been accompanied by the proliferation of a large number of fake pharmacy websites (i.e., ones selling fake drugs, providing fictitious information, and/or failing to ship the agreed upon goods altogether). According to a Food and Drug Administration study, less than 10% of the 12,000 Internet pharmacies examined were legitimate (Krebs, 2005). The number of users visiting fake online pharmacies has also tripled over the past year (Greenberg, 2008).

Fake pharmacy websites are likely to offer a distinct set of challenges to standard fake website detection methods due to a large number of internal links within a site. Though such internal links are typical within an online store (like a pharmacy), they are not common in escrow, bank, and other e-commerce sites examined by other spoof site detection studies (Abbasi and Chen, 2007; 2009a; 2009b; Abbasi et al., 2008; 2010; Liu et al., 2006). This provides the motivation to study the performance of state-of-the-art algorithms on pharmacy sites. In addition, since fake pharmacy website developers spend millions of dollars on black-hat search engine optimization in order to make their websites more visible (Greenberg, 2008) we contend that they may be susceptible to detection by link-based techniques designed to combat web spam (Gyongyi and Garcia-Molina, 2005). However, it is unclear how effective web spam detection methods, which make specific assumptions about the linkage between good and bad pages, would be on fake pharmacy websites.

## 2. Existing Link-Based Techniques

Several link-based techniques have been proposed for detecting web spam. Table 1 presents a summary of select link-based algorithms used in prior fake website detection research. These techniques can be categorized according to their propagation mechanisms, which are based on the assumptions they make regarding how real and fake web pages are connected to one another. Single class propagation algorithms begin with a seed set of known good or bad pages. These pages' 'goodness' or 'badness' is then propagated through the graph via their inlinks and/or outlinks. Examples of single class propagation algorithms include BadRank, TrustRank, AntiTrustRank, and ParentPenalty (Gyongyi et al., 2004; Krishnan and Raj, 2006; Wu and Davison, 2005). Dual class propagation algorithms, such as QoC and QoL, utilize good and bad seed pages and also propagate scores through both inlinks and outlinks (Zhang et al., 2006). A third class of algorithms like PageRank does not use any seed pages; it is a purely unsupervised scoring algorithm that was not designed to detect bad pages (Page et al., 1998). Nevertheless, it is effective at detecting fake web pages that do not utilize link farms and is therefore often used as a baseline (Zhang et al., 2006).

However, it is unclear as to how these link-based algorithms designed for web-spam detection (Gyongyi and Garcia-Molina, 2005; Gyongyi et al., 2004; Krishnan and Raj, 2006; Wu and Davison, 2005) would perform on fake pharmacy websites. In addition, there is a gap in the understanding of the effect of inlinks and outlinks and node granularity (webpage vs. website) on these algorithms. In this study, we explore the following research questions to address these issues:

- How effective are existing link-based algorithms for the detection of fake online pharmacy websites?
- What impact does the propagation mechanism, including the number of classes propagated and whether the algorithm uses inlinks and/or outlinks, have on performance?
- Which granularity (page or site) is the most suited in this domain?

**Table 1:** Selected Link-based Techniques

| Algorithm | Seed Pages | Propagation |
|---|---|---|
| BadRank (mentioned in Wu and Davison, 2005) | $E(A) = 1$, for known bad pages<br>$E(A) = 0$, for all other pages | $BR(A) = E(A)(1-d) + d\sum_{i=1}^{n} \frac{BR(T_i)}{C(T_i)}$<br>where:<br>$E(A)$ is the initial BadRank score of A<br>$BR(T_i)$ is the BadRank score of $T_i$, an outlink of A<br>$C(T_i)$ is the number of inlinks for $T_i$<br>$d$ is a tunable parameter between 0 and 1 |
| TrustRank (Gyongyi et al., 2004) | $E(A) = 1$, for known good pages<br>$E(A) = 0$, for all other pages | $TR(A) = \frac{E(A)(1-d)}{n} + d\sum_{i=1}^{n} \frac{TR(I_i)}{C(I_i)}$<br>where:<br>$E(A)$ is the initial TrustRank score of A<br>$TR(I_i)$ is the TrustRank score of $I_i$, an inlink of A<br>$C(I_i)$ is the number of outlinks for $T_i$<br>$d$ is a tunable parameter between 0 and 1 |
| PageRank (Page et al., 1998) | Not applicable | $PR(A) = \frac{(1-d)}{n} + d\sum_{i=1}^{n} \frac{PR(I_i)}{C(I_i)}$<br>where:<br>$PR(I_i)$ is the PageRank score of $I_i$, an inlink of A<br>$C(I_i)$ is the number of outlinks for $T_i$<br>$d$ is a tunable parameter between 0 and 1 |
| AntiTrustRank (Krishnan and Raj, 2006) | $E(A) = 1$, for known bad pages <br> $E(A) = 0$, for all other pages | $AR(A) = \dfrac{E(A)(1-d)}{n} + d\sum_{i=1}^{n}\dfrac{AR(I_i)}{C(I_i)}$ <br> where: <br> $E(A)$ is the initial AntiTrustRank score of $A$ <br> $AR(I_i)$ is the AntiTrustRank score of $I_i$, an inlink of $A$ <br> $C(I_i)$ is the number of outlinks for $T_i$ <br> $d$ is a tunable parameter between 0 and 1 |
| ParentPenalty (Wu and Davison, 2005) | $S(A) = 1$, if the number of common sites in A's inlinks and outlinks exceeds a threshold $t$ <br> $S(A) = 0$, for all other pages | $S(A) = 1$, if $\sum_{i=1}^{n} S(T_i) \geq p$ <br> $S(A) = 0$, otherwise <br> where: <br> $S(T_i)$ is the score of $T_i$, an outlink of $A$ <br> $p$ is a threshold parameter |
| QoL (Zhang et al., 2006) | $EL(A) = 1$, if the number of good pages A points to is greater than the number of bad pages, and the number of bad pages is less than the threshold parameter $k$ <br> $EL(A) = -1$, for all other pages | $QoL(A) = EL(A)(1-d) + d\left(\sum_{i=1}^{n}\left(\beta\dfrac{QoC(T_i)}{C(T_i)} + (1-\beta)\dfrac{QoL(T_i)}{C(T_i)}\right)\right)$ <br> where: <br> $EL(A)$ is the initial score of $A$ <br> $QoC(T_i)$ is the QoC score of $T_i$, an outlink of $A$ <br> $QoL(T_i)$ is the QoL score of $T_i$ <br> $C(T_i)$ is the number of inlinks of $I_i$ <br> $\beta$ and $d$ are tunable parameters between 0 and 1 |
| QoC (Zhang et al., 2006) | $EC(A) = 1$, for known good pages <br> $EC(A) = 0$, for all other pages | $QoC(A) = EC(A)(1-d) + d\left(\sum_{i=1}^{n}\left(\alpha\dfrac{QoC(I_i)}{C(I_i)} + (1-\alpha)\dfrac{QoL(I_i)}{C(I_i)}\right)\right)$ <br> where: <br> $EC(A)$ is the initial score of $A$ <br> $QoC(I_i)$ is the QoC score of $I_i$, an inlink of $A$ <br> $QoL(I_i)$ is the QoL score of $I_i$ <br> $C(I_i)$ is the number of outlinks of $I_i$ <br> $\alpha$ and $d$ are tunable parameters between 0 and 1 |

### 3. Research Testbed and Design

We collected 150 legitimate and 150 fraudulent online pharmacy websites. The URLs for these websites were obtained from reputable sources including the National Association of Boards of Pharmacy (www.nabp.net) and LegitScript (www.legitscript.com). All pages within each website were collected leading to an initial collection of approximately 1.2 million pages. Surprisingly, the fake pharmacy websites were much larger than the legitimate ones, with nearly twice as many pages. The entire hyperlink graph for these 300 seed URLs was collected over a 3-month period using link expansion. This involved iteratively collecting all inlinks and outlinks of queue pages, adding the newly collected pages to the queue, and repeating the process (where the initial queue contained all pages from the 300 seed websites). The graph encompassed over 15.5 million pages from approximately 930,000 websites/domains, with 81 million unique links.

We evaluated seven common algorithms for fake-website detection using our dataset - BadRank, TrustRank, PageRank, AntiTrustRank, ParentPenalty, QoL, and QoC. A bootstrapping approach was employed where the 300 websites were randomly split into 150 training and 150 test cases for 30 independent runs. For any particular run, the algorithms were only evaluated on the pages associated with the 150 test websites. Each algorithm's parameters were tuned extensively; the settings yielding the best results on the testing data were utilized. Consistent with prior research (Wu and Davison, 2005), the evaluation metrics employed included overall accuracy and class-level precision, recall, and F-measure. In order to answer the third research question, we conducted two independent experiments using the above methodology – one with page level and the other with site level data.

## 4. Experimental Results

In the first experiment, the algorithms were run on the page-level graph. The results, averaged across the 30 bootstrap runs, are presented in Table 2. Both, page-level and site-level results were computed. The site-level results were aggregated, where a website was considered legitimate only if the majority of its pages were deemed legitimate by the algorithm. QoC had the best overall performance, followed by QoL. Single class propagation algorithms such as BadRank, AntiTrustRank, and ParentPenalty had higher performance on legitimate websites but were highly ineffective on fraudulent ones (with low fake website recall). Their inability to effectively propagate badness scores to fake pages resulted in many false positives (i.e., fraudulent pages classified as being legitimate). Similarly, TrustRank performed poorly on legitimate websites since it it could not effectively propagate goodness scores to many of the legitimate websites. However, it performed well on the fraudulent websites, resulting in high overall page-level accuracy (since the fraudulent websites comprised a much larger share of the web pages in the test bed).

**Table 2:** Site-Level Classification Performance using Page-level Graph

| Site Level Performance | | | | | | | |
|---|---|---|---|---|---|---|---|
| Algorithm | Overall Accuracy | Legitimate Websites | | | Fake Websites | | |
| | | F-Meas. | Precision | Recall | F-Meas. | Precision | Recall |
| BadRank | 52.29 | 67.67 | 51.18 | 99.87 | 8.92 | 96.56 | 4.71 |
| TrustRank | 59.62 | 38.86 | 79.59 | 25.78 | 69.84 | 55.76 | 93.47 |
| PageRank | 60.29 | 67.09 | 57.32 | 80.93 | 49.85 | 67.47 | 39.64 |
| AntiTrustRank | 57.36 | 69.00 | 54.22 | 94.93 | 31.50 | 80.22 | 19.78 |
| ParentPenalty | 51.07 | 67.15 | 50.54 | 100.00 | 4.61 | 100.00 | 2.13 |
| QoL | 74.51 | 66.25 | 97.70 | 50.22 | 79.51 | 66.54 | 98.80 |
| QoC | 88.09 | 86.93 | 96.14 | 79.42 | 89.05 | 82.54 | 96.76 |

| Page Level Performance | | | | | | | |
|---|---|---|---|---|---|---|---|
| Algorithm | Overall Accuracy | Legitimate Websites | | | Fake Websites | | |
| | | F-Meas. | Precision | Recall | F-Meas. | Precision | Recall |
| BadRank | 27.90 | 41.78 | 27.04 | 99.98 | 3.61 | 99.32 | 1.90 |
| TrustRank | 83.21 | 54.46 | 88.95 | 40.28 | 89.32 | 82.00 | 98.46 |
| PageRank | 41.13 | 35.85 | 26.62 | 63.13 | 44.98 | 71.68 | 34.21 |
| AntiTrustRank | 31.65 | 37.08 | 24.90 | 78.97 | 22.56 | 64.13 | 14.61 |
| ParentPenalty | 33.13 | 43.36 | 28.27 | 99.58 | 15.23 | 89.24 | 8.69 |
| QoL | 84.73 | 60.23 | 94.39 | 45.19 | 90.22 | 83.10 | 99.12 |
| QoC | 94.11 | 87.54 | 94.79 | 81.61 | 95.94 | 93.60 | 98.47 |

Figure 1 shows the accuracies on the pages ranked in the top 100,000 (good and bad) for each algorithm, based on the algorithm's scores. For instance, since BadRank assigns badness scores, the top 100,000 good pages using BadRank would be the ones with the lowest scores (while the top 100,000 bad pages would be ones with the highest scores). The figure reaffirms the findings from Table 2, with the performance of QoC and QoL near the top in both good and bad pages, while the other algorithms performing well on only one of the classes. Both QoC and QoL are dual class propagation algorithms and they are also the only ones that incorporate inlink and outlink information during the propagation process. Pair-wise t-tests conducted on the 30 bootstrap runs revealed that these two techniques significantly outperformed the other 5 single class propagation algorithms (all 10 p-values < 0.001, n=30). Moreover, QoC also significantly outperformed QoL (p-value < 0.001, n=30).

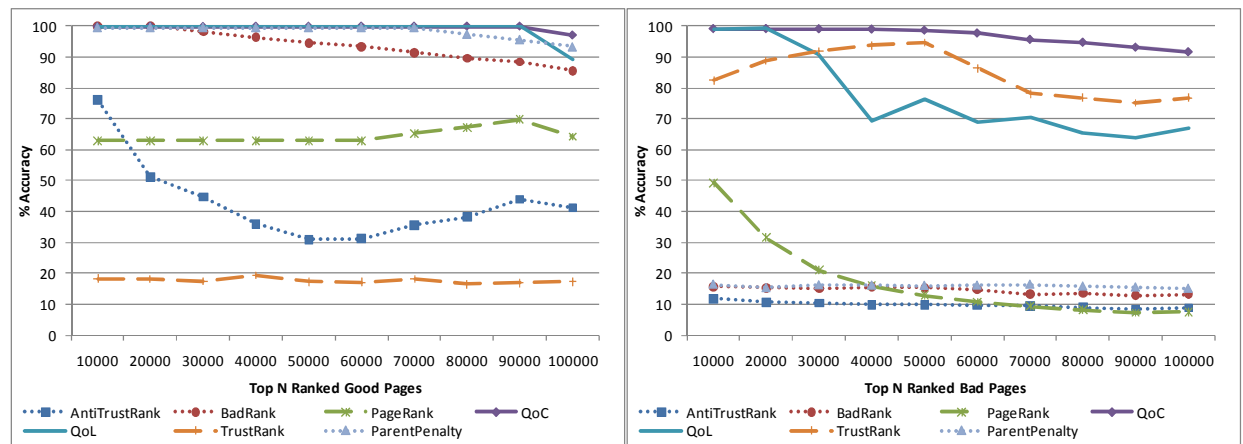

**Figure 1:** Accuracy on Top 100,000 Ranked (a) Good and (b) Bad Pages

In the second experiment, the seven algorithms were run on the site-level graph. This graph contained a single node for each website and only inter-site links, resulting in a smaller graph. The results, averaged across the 30 bootstrap runs, are presented in Table 3. Once again, QoC and QoL had the best performance, attaining over 90% overall accuracy. In general, the results using site-level graphs were better. Five of the seven algorithms attained better results, with improvements ranging from 7% to 21%. Pair-wise t-tests conducted on accuracy were significant for these five algorithms (all five p-values <0.001, n=30). This suggested that internal links within a site may play an important role in this domain. However, PageRank and ParentPenalty's performance fell, with the decrease significant for PageRank. We are in the process of examining the results closely to better understand the performance of these algorithms.

**Table 3:** Classification Performance using Site-level Graph

| Algorithm | Overall Accuracy | Legitimate Websites | | | Fake Websites | | |
|---|---|---|---|---|---|---|---|
| | | F-Meas. | Precision | Recall | F-Meas. | Precision | Recall |
| BadRank | 65.22 | 73.02 | 59.66 | 94.13 | 51.01 | 86.39 | 36.31 |
| TrustRank | 81.67 | 82.11 | 80.15 | 84.49 | 81.12 | 83.93 | 78.84 |
| PageRank | 53.89 | 68.30 | 52.04 | 99.33 | 15.40 | 92.81 | 8.44 |
| AntiTrustRank | 70.44 | 75.93 | 64.09 | 93.20 | 61.66 | 87.70 | 47.69 |
| ParentPenalty | 50.67 | 53.56 | 48.13 | 60.44 | 40.45 | 46.88 | 34.89 |
| QoL | 92.84 | 92.57 | 95.69 | 89.82 | 93.08 | 90.59 | 95.87 |
| QoC | 95.71 | 95.62 | 97.43 | 93.96 | 95.80 | 94.26 | 97.47 |

## 5. Conclusions and Future Directions

In this preliminary study, we investigated the effectiveness of several existing link-based algorithms for the detection of fake online pharmacy websites. The results revealed that algorithm and graph characteristics both play an important role in the accurate detection of fake pharmacy websites. The dual propagation algorithms that considered inlink and outlink information for good and bad seed pages provided the best performance. Further, site-level graphs yielded considerably better results than page-level ones for most algorithms evaluated.

Based on these findings, we have identified some future research directions. We intend to bolster the algorithm comparison analysis by incorporating receiver operating characteristic (ROC) curves showing the impact of different page and site-level thresholds on true and false positive rates. Given the large imbalance between the number of legitimate and fake pharmacy websites in existence, we also intend to explore the impact of using different ratios of legitimate and fake sites in the training data. Given the strengths and weaknesses of these algorithms, we are also exploring various fusion strategies for combining the algorithms in a manner that harnesses their collective discriminatory potential. These include using them as features in machine learning classification algorithms (e.g., Decision trees, Logit regression) or using the output of certain single class propagation algorithms with high class-level precision as input for others. We are also evaluating the efficacy of hybrid classification approaches that combine content analysis with link-based algorithms.

**References**

Abbasi, A. and Chen, H. "Detecting Fake Escrow Websites using Rich Fraud Cues and Kernel Based Methods," In the 17th Annual Workshop on Information Technologies and Systems, Montreal, Canada, December 8-9, 2007, pp. 55-60.

Abbasi, A., Zhang, Z., and Chen, H. "A Statistical Learning Based System for Fake Website Detection," In the Workshop on Secure Knowledge Management, Dallas, Texas, November 3-4, 2008.

Abbasi, A. and Chen, H. "A Comparison of Fraud Cues and Classification Methods for Fake Escrow Website Detection," Information Technology and Management, 10(2), 2009a, pp. 83-101.

Abbasi, A. and Chen, H. "A Comparison of Tools for Detecting Fake Websites," IEEE Computer, 42(10), 2009b, pp. 78-86.

Abbasi, A., Zhang, Z., Zimbra, D., Chen, H., and Nunamaker Jr., J. F. "Detecting Fake Websites: The Contribution of Statistical Learning Theory," MIS Quarterly, 2010, forthcoming.

Easton, G. "Clicking for Pills," British Medical Journal, 334(7583), 2007, pp. 14-15.

Greenberg, A. "Pharma's Black Market Boom," Forbes.com, August 26, 2008.

Gyongyi, Z. and Garcia-Molina, H. "Spam: It's not Just for Inboxes Anymore," IEEE Computer, 38(10), 2005, pp. 28-34.

Gyongyi, Z., Garcia-Molina, H., and Pedersen, J. "Combating Web Spam with Trust Rank," In Proceedings of the 13th International Conference on Very Large Data Bases, 2004, pp. 576-587.

Krebs, B. "Few Online 'Canadian Pharmacies' Based in Canada, FDA Says," WashingtonPost.com, June 14, 2005.

Krishnan, V. and Raj, R. "Web Spam Detection with Anti-Trust Rank," In Proceedings of the 2nd International Workshop on Adversarial Information Retrieval on the Web, 2006, pp. 37-40.

Liu, W., Deng, X., Huang, G., and Fu, A. Y. "An Antiphishing Strategy Based on Visual Similarity Assessment," IEEE Internet Computing, 10(2), 2006, pp. 58-65.

Page, B., Brin, S., Motwani, R., and Winograd, T. "The PageRank Citation Ranking: Bringing Order to the Web," Stanford University Technical Report, 1998.

Wilford, B. B., Smith, D. E., and Bucher, R. "Prescription Stimulant Sales on the Internet," Pediatric Annals, 35(8), 2006, pp. 575-578.

Wu, B. and Davison, B. D. "Identifying Link Farm Spam Pages," In Proceedings of the 14th International World Wide Web Conference, 2005, pp. 820-829.

Zhang, L., Zhang, Y., and Zhang, Y., and Li, X. "Exploring Both Content and Link Quality for Anti-Spamming," In Proceedings of the 6th IEEE International Conference on Computer and Information Technology, 2006, pp. 37-42.